\documentclass[aps,preprint,showpacs]{revtex4}

\usepackage{graphicx,epsfig}
\usepackage{amssymb}
\begin{document}

\title{Quasinormal modes of nonlinear electromagnetic black holes from unstable null geodesics.}

\author{N. Bret\'on$^{1}$}

\author{L. A. L\'opez $^{2}$}

\affiliation{$^{1}$ Dpto de F\'isica, Centro de Investigaci\'on y de Estudios Avanzados \\
del I.P.N., Apdo. 14-740, D.F., M\'exico.}

\affiliation{$^{2}$ \'Area Acad\'emica de Matem\'aticas y F\'isica, UAEH, \\
Carretera Pachuca-Tulancingo Km. 4.5, C. P. 42184, Pachuca, M\'exico.}

\begin{abstract}

The expressions for the quasinormal modes
(QNMs)  of black holes with nonlinear electrodynamics, calculated in
the eikonal approximation, are presented.
In the eikonal limit QNMs of black holes are
determined by the parameters of the circular null geodesics.
The unstable circular null orbits are derived from the effective metric
that is the one obeyed by  light rays under the influence of a
nonlinear electromagnetic field. As an illustration we calculate the QNMs of four nonlinear electromagnetic black holes, two singular and two regular, namely from Euler-Heisenberg and Born-Infeld theories, for singular, and  the magnetic Bardeen black hole and the one derived by  Bronnikov for regular ones.  Comparison is shown with the QNMs of the linear electromagnetic counterpart, their Reissner-Nordstr\"{o}m black hole.

\end{abstract}


\maketitle

\section{Introduction}

When a black hole undergoes perturbations, the resulting behavior can be described in three stages. The first
stage corresponds to radiation due to the initial conditions of the perturbations. The second stage corresponds to
damped oscillations with complex frequencies. The third stage in general corresponds to a power law decay of the fields. The modes of such oscillations are called quasi-normal modes (QNMs). The frequencies of QNMs are independent of initial perturbations, they are then the intrinsic imprint of the response of the black hole to external perturbations.

Besides the importance of QNMs  in the analysis of black hole stability, they play an outmost role in characterising  gravitational wave signals, as the ones recently detected at LIGO, as well as the promised ones jointly with VIRGO collaboration and the planned space antenna project  LISA \cite{Abbott:2016blz}.
QN resonances, being the characteristic ``sound"  of the black hole itself,  are crucial to identify the spacetime parameters, especially the mass and angular momentum of the black hole, but as well are going to be important on identifying additional physical parameters arising from more realistic black hole models.

There are also theoretical reasons that justify the study of QNMs,  one of them from Loop Quantum Gravity (LQG). It has been observed that for asymptotically flat black holes, the real part of the high overtones of QNMs coincide with the Barbero-Immirzi parameter \cite{Kunstatter:2002pj}; this parameter measures the size of the quantum area in Planck units in relation to the counting of microstates in LQG.

Mashhoon \cite{Mashhoon:1985cya} \cite{Ferrari:1984zz} has suggested an analytical technique of calculating the QNMs in the geometric-optics (eikonal) limit. The basic idea is to interpret the black-hole free oscillations in terms of null particles trapped at the unstable circular orbit and slowly leaking out.
The real part of the complex quasinormal frequencies (QNFs) is determined by the angular velocity at the unstable null geodesic; the imaginary part is related to the instability time scale of the orbit.  In such a way that QNMs can be determined from the unstable null geodesics that are the orbits attached to the maximum of the effective potential barrier felt by light rays on their interaction with the black hole. In this sense Cardoso \cite{Cardoso:2008bp} showed  the relationship among unstable null geodesics, Lyapunov exponents and quasinormal modes in a stationary spherically symmetric space-time.

It is  well known  that in situations involving strong electromagnetic fields, the linear superposition does not hold and nonlinear effects, for instance the creation of electron-positron pairs or scattering of light by light, are very likely to occur. These situations are described by Quantum Electrodynamics (QED); alternatively, classical theories that include these nonlinear phenomena in an effective way may be useful. Among these theories are the Euler-Heisenberg (EH) and the Born-Infeld (BI) \cite{Born:1934gh}. Moreover, nonlinear electromagnetic (NLEM) theories have attracted attention lately
due to their ability in suppressing the singularity in some black hole solutions.
As a result of the nonlinear interaction, light rays do not follow the null geodesics of the background metric, but do follow the null geodesics of an effective metric that depends on the nonlinear electromagnetic energy momentum tensor \cite{Novello:1999pg} \cite{Gutierrez:1981ed}.

In \cite{Fernando:2012yw} applying the ideas of Cardoso, were determined the QNMs of the regular magnetic black hole model proposed by  Bardeen, which is a solution of nonlinear electrodynamics coupled to Einstein gravity. QNMs of nonlinear electromagnetic black holes  were  computed, using the  WKB method, for  the Born-Infeld black hole in \cite{Fernando:2004pc} and for the  Bronnikov black hole in \cite{Li:2014fka}.
Our study differs from previous ones in the fact that we use the effective metric to determine the unstable circular orbits followed by light rays. Otherwise the QNFs correspond to massless test particles, that indeed follow, in the eikonal approximation, the null geodesics of the background metric. Those QNFs do not correspond to light rays, that in the presence of strong electromagnetic fields do interact among themselves giving rise to nonlinear effects.

Using the effective metric and the corresponding effective potential  we derive the Lyapunov exponent expression, that is related to the imaginary part of the quasinormal frequencies (QNF); as an illustration we address two examples of singular black holes: the Born-Infeld and a magnetic Euler-Heisenberg one, as well as two examples of regular black holes, the Bardeen model for a magnetic self-gravitating black hole and one solution derived by Bronnikov.

The paper is organized as follows:
In  Sect. II we give a brief explanation to determining  the QNFs using the unstable null geodesics and the Lyapunov exponent. In Sect. III a short
summary of the nonlinear electrodynamics for a static spherically symmetric spacetime is presented as well as the effective NLEM metric. In Sect. IV
the expressions for the real and imaginary parts of the QNFs in terms of the NLEM Lagrangian are given. In Sect. V we analyze  the QNFs of the four  examples above mentioned. In each case the QNFs are compared with the ones corresponding to the massless test particles and light rays of the Reissner-Nordstrom (RN) black hole.  Conclusions are given in the last section.

\section{QNM and  the Lyapunov exponent}

The connection between the QNMs and bound states of the the inverted black hole effective potential was pointed out in
\cite{Ferrari:1984zz}.
In \cite{Cardoso:2008bp}  it is shown that, in the eikonal limit,  the QNMs of black holes in any dimensions are determined by the parameters of the circular null geodesics. The real part of the complex QNM frequencies is determined by the angular velocity at the unstable null geodesics. The imaginary part is related to the instability time scale of the orbit, and therefore related to the Lyapunov exponent that is its inverse. Lyapunov exponents are a measurement  of the average rate at which nearby trajectories converge or diverge in the phase
space. A positive Lyapunov exponent indicates a divergence between nearby trajectories, i.e., a high sensitivity to
initial conditions.  In the case of stationary, spherically symmetric spacetimes  it turns out that this exponent can be expressed as the second derivative of the effective potential evaluated at the radius of the unstable circular null orbit.
It was also shown the  agreement of the so calculated QNMs with the analytic WKB approximation,

\begin{equation}\label{QNM}
\omega_{QNM}=\Omega_{c}l - i (n+\frac{1}{2})| \lambda|
\end{equation}
where $n$ is the overtone number and $l$ is the angular momentum of the perturbation.   $\Omega_{c}$ is the angular velocity at the unstable null geodesic and  $\lambda$ is the Lyapunov exponent, determining the instability time scale of the orbit. From the equations of motion for a test particle in the static spherically symmetric  (SSS) spacetime, $\dot{r}^{2}=V_{r}$, where $V_{r}$  is the effective potential for radial motion, circular geodesics are determined from the conditions $V(r_{c})=V^{'}(r_{c})=0$ where $r_c$ is the radius of the circular orbit.
The  Lyapunov exponent in terms of the second derivative of the effective potential is given by

\begin{equation}\label{expLyapunov}
\lambda=\sqrt{\frac{V_{r}^{''}}{2\dot{t}^{2}}},
\end{equation}
where $t$ is the time coordinate. The dot denotes the derivative with respect to an affine parameter and the prime
stands for the derivative with respect to  $r$ .
The orbital angular velocity is given by

\begin{equation}\label{angular}
\Omega_{c}=\frac{d{\varphi}}{d{t}}=\frac{\dot{\varphi}}{\dot{t}}.
\end{equation}

For our purpose both expressions should be evaluated at $r_c$, the radius of the unstable null circular orbit, that is the orbit with an impact parameter ${b_c}$  related to the effective potential as $V_r=1/{b_c}^2$, i. e. orbits with an energy equal to the maximum of the effective potential.

For a static  spherically symmetric background

\begin{equation}\label{sss}
ds^{2}=f(r)dt^{2}-\frac{1}{g(r)}dr^{2}-r^{2}d\Omega^{2},
\end{equation}
the energy  $E$   and the angular momentum $L$ of a test particle  are conserved quantities,

\begin{equation}\label{constan}
f(r)\dot{t}= - E= {\rm const},\;\;\;\;\;\;\;\ r^{2}\dot{\varphi}= - L= {\rm const}.
\end{equation}

For equatorial orbits ($\theta= \pi/2$), the equation for radial motion is $\dot{r}^2=V_r$. For the case of a static spacetime $f=g$ in  (\ref{sss}), and  the effective potential is  given by

\begin{equation}
V_{r}=g(r)\left[ \frac{E^{2}}{f(r)}-\frac{L^{2}}{r^{2}}\right]  = E^2- f \frac{L^{2}}{r^{2}}.
\end{equation}

Then the Lyapunov exponent, related to the imaginary part of the QNF,  from (\ref{expLyapunov}) is given by

\begin{equation}\label{Lyapunov}
\lambda^{2}=\frac{f_{c}}{2r_{c}^{2}}\left[2f_{c}-r_{c}^{2}f^{''}_{c}\right],
\end{equation}

while the orbital angular velocity, that is proportional to the QNF real part,  is given by

\begin{equation}\label{angular1}
\Omega_{c}=\frac{L}{r_c^2} \frac{f_c}{E}= \sqrt{\frac{f_c}{r_c^2}}.
\end{equation}
In the previous expressions (\ref{Lyapunov}) and (\ref{angular1}) the conditions for a circular orbit, $V_r=0$ and $V_r'=0$ that amount, respectively,  to

\begin{equation}\label{nullconds}
\frac{E^2}{L^2}= \frac{f}{r^2}, \quad {\rm and} \quad 2f-rf'=0,
\end{equation}
have been incorporated.

In the following  section (\ref{Lyapunov}) and (\ref{angular1}) will be determined from the effective metric, and $\lambda$ and $\Omega_c$ will be written in terms of the NLEM Lagrangian.

\section{QNMs of NLEM black holes from the effective metric}

The action for  gravitation  coupled to  and electromagnetic field is given by

\begin{equation}\label{action}
S=\frac{1}{16\pi}\int d^{4}x\sqrt{-g}[R-L(F)],
\end{equation}
where $R$ is the scalar curvature and $L$ is an arbitrary function of the electromagnetic invariant  $F=F^{\mu\nu}F_{\mu\nu}$,  with
$F_{\mu\nu}=\partial_{\mu}A_{\nu}-\partial_{\nu}A_{\mu}$ being  the electromagnetic field tensor. For the Maxwell theory the Lagrangian is directly proportional to $F$, $ L_{M}=F$. We note that the most general Lagrangian can depend on both electromagnetic invariants, $F$ and  $G={\cal F}^{\mu\nu}F_{\mu\nu}$, where  ${\cal F}^{\mu\nu}$ is the dual field-strength electromagnetic tensor.  In this work we shall restrict ourselves to lagrangians depending only on $F$ but otherwise completely general. Roughly speaking this restriction  has the consequence of having solutions with only electric or magnetic charges, but not both.

The electromagnetic tensor compatible with spherical symmetry has two nonzero components,
$F_{01} = -F_{10}$ and $F_{23} = -F_{32}$, corresponding to the radial electric and magnetic fields, with

\begin{equation}
r^{2}L_{F}F^{10}=q_{e}, \quad  F_{23}=q_{m}\sin \theta,
\end{equation}

where $q_{e}$ and $q_{m}$ are the electric and magnetic charges, respectively, while the subindex $F$ means derivative with respect to $F$. Adopting  the definitions

\begin{equation}
f_{e}= 2F_{01}F^{10}=2q_{e}^{2}L_{F}^{-2}r^{-4}\geq 0 , \;\;\;\;\;\;\ f_{m}= 2F_{23}F^{23}=2q_{m}^{2}r^{-4}\geq 0
\end{equation}

$F=f_m-f_e$ and the energy-momentum tensor can be written as

\begin{equation}
T^{\mu}_{\nu}=\frac{1}{2}{\rm  diag}
(L + 2 f_e \:L_F,L + 2 f_e \:L_F, L - 2 f_m\: L_F,L - 2 f_m \:L_F),
\end{equation}

The nonlinearity of the electromagnetic field modifies  light trajectories that regularly are the null geodesics of the background metric  $g_{\mu\nu}$.
In nonlinear electromagnetism,  instead, photons do propagate along null geodesics of an effective
geometry with metric tensor $\gamma^{\mu\nu}_{\rm eff}$ that depends on the nonlinear theory.
The discontinuities of the electromagnetic field propagate obeying the equation for the characteristic
surfaces.  Then the effective metric can be derived from the analysis of the
characteristic surfaces  \cite{Novello:1999pg}  \cite{Gutierrez:1981ed} . The corresponding equation  for the gradient of the characteristic surfaces $S_{, \mu}$ is

\begin{equation}\label{null}
(L_{F}g^{\mu\nu}-4L_{FF}F^{\mu}_{\alpha}F^{\alpha\nu})S_{, \mu}S_{, \nu}=\gamma^{\mu\nu}_{\rm eff}S_{, \mu}S_{, \nu}=0.
\end{equation}

Calculating the metric components $\gamma^{\mu\nu}_{\rm eff}$  the effective metric of the SSS spacetime is given by

\begin{equation}\label{efec}
ds_{\rm eff}^{2}=(L_{F}G_{m})^{-1}\left\{G_{m}G_{e}^{-1}\left( f(r)dt^{2} - \frac{1}{f(r)}dr^{2}\right)-r^{2}d\Omega^{2}\right\}.
\end{equation}

$G_m$ and $G_e$, the magnetic and electric factors that make the difference between the linear and nonlinear electromagnetism, are given by

\begin{equation}\label{Gs}
G_{m}=\left(1+4L_{FF}\frac{q_{m}^{2}}{L_{F}r^{4}}\right) , \;\;\;\;\;\;\ G_{e}=\left(1-4L_{FF}\frac{q_{e}^{2}}{L_{F}^{3}r^{4}}\right);
\end{equation}
and in the linear limit become equal to one.
In determining null geodesics conformal factors can be ignored, since they do not modify the null geodesics.  Considering the geodesic motion in the equatorial plane of the  effective spacetime (\ref{efec}), not including the conformal factor  $(L_{F}G_{m})^{-1}$, the corresponding  effective  potential $V_{r}$   is

\begin{equation}
V_{r}= G_{m}^{-1}G_{e}\left[G_{m}^{-1}G_{e} E^2 -\frac{f(r){L}^{2}}{r^{2}}\right].
\end{equation}

When we apply the first  conditions for the (unstable) null circular orbits  $V(r_{c})=0$ , we obtain

\begin{equation}
\frac{{E}^{2}}{{L}^{2}}=\left(\frac{G_{m}}{G_{e}}\right)_{r_{c}}\frac{f_{c}}{r_{c}^{2}}
\end{equation}

and jointly with  the second condition, $V^{'}(r_{c})=0$, the radius $r_c$ of the circular null orbit is given by one of the roots of  the following equation,

\begin{equation}\label{rcEq.}
\left( \frac{G_{e}}{G_{m}}\right) \left(\frac{ f_{c}^{'}}{ f_{c}}- \frac{2}{r_{c}}\right) -\left( \frac{G_{e}}{G_{m}}\right)^{'}_{r_{c}}=0.
\end{equation}
Such root should be greater than the horizon radius, $r_c>r_h$;
subscript  ``c"  means that the quantity in question is evaluated at the radius $r= r_{c}$. This radius is also known as the radius of the photosphere, and defines the sphere of unstable circular photon trajectories.

The  Lyapunov  exponent (\ref{expLyapunov}) for the effective metric takes the form;

\begin{equation}\label{expLyapunov1}
\lambda^{2}=\frac{f_{c}r^{2}_{c}}{2}\left[  \frac{f_{c}}{r_{c}^{2}} \frac{G_{m}}{G_{e}} \left( \frac{G_{e}}{G_{m}}\right)^{''}_{c}- \left(\frac{f}{r^{2}}\right)^{''}_{c}\right]
\end{equation}

while the angular velocity (\ref{angular}) changes to
\begin{equation}\label{angular2}
\Omega_{c}= \sqrt{\frac{G_{m}}{G_{e}} \frac{f_{c}}{r_{c }^2}}.
\end{equation}

Equations (\ref{expLyapunov1}) and (\ref{angular2}),  the nonlinear electromagnetic version of  (\ref{Lyapunov}) and (\ref{angular1}),  determine the quasinormal frequencies, imaginary and real parts, respectively,  for  NLEM black holes in the  eikonal approximation.

From the  expressions  (\ref{Gs}) and having determined the sign of $L_{FF}/L_{F}$ it can be defined if $G_{m}$ and $G_{e}$  are greater or less than one.
From that knowledge and the expression of $\Omega_c$ we can assert if the real part of the QNF is enhanced or suppressed as compared to the linear counterpart. For instance if  $L_{FF}/L_{F }<0$ then $G_{m} \le 1$ and $G_{e} >1$ and consequently, from (\ref{angular2}),  $\Omega_c$  gets a smaller value, resulting in a suppression of the real part of QNF $\omega_r$.

The analysis is not so straightforward for the imaginary part, eq. (\ref{expLyapunov1}), since it is not obvious which the sign of the second derivative of $G_{e}/G_{m}$ will be. In this case we need more specific information about the NLEM lagrangian.

In the next section examples of different NLEM black holes are given and the QNF  due to the NLEM field are compared with those corresponding to massless test particles as well as with the QNF originated from the  linear electromagnetic black hole, the SSS solution to the Einstein-Maxwell equations, the Reissner-Nordstrom.

\section{Examples}
We calculate the QNFs of four SSS black hole solutions corresponding to different NLEM theories coupled to gravity.   First will be addressed two examples of singular NLEM black holes, the Born-Infeld (BI) and a magnetic solution for the Euler-Heisenberg electrodynamics coupled to gravity.  Then we analyse the QNM response  of two regular NLEM black holes: the Bardeen magnetic monopole and one Bronnikov solution, comparison with massless particles and linear electromagnetism is established as well. It is considered only the fundamental frequency $n=0$, that is the least damped mode.

\subsection{The QNM of the Reissner-Nordstrom black hole}

The RN is the SSS solution to the action (\ref{action}) with $L=F$, $L_{F}=1$ and $L_{FF}=0$ and then the nonlinear factors are $G_e=1$ and $G_m=1$. For each example addressed in what follows RN will be the reference to analize the QNFs behavior. Briefly the  RN case is exposed.

In the eikonal or geometric-optics limit, the QNFs are given by (\ref{QNM}) with
$\lambda$ and $\Omega_{c}$ calculated as in   (\ref{Lyapunov}) and (\ref{angular1}), respectively. For the RN black hole, the metric function in the line element  (\ref{sss}) is given by

\begin{equation}
f(r)=g(r)= 1- \frac{2M}{r}+ \frac{Q^2}{r^2},
\end{equation}

while the circular null orbit radius $r_c$  is calculated from (\ref{nullconds}), that in the RN case amount to the quadratic polynomial

\begin{equation}\label{rcRN}
r_c^2-3Mr_c+2Q^2=0, \quad r_c=  \frac{1}{2} (3 + \sqrt{9 - 8 Q^2}).
\end{equation}
where we have used a dimensionless coordinate  $r \to r/M$ and $Q\to Q/M$. The functions $\lambda$ and $\Omega_c$ are given by
(\ref{Lyapunov}) and (\ref{angular1}),

\begin{equation}
\lambda^2= \frac{1}{r^{6}_{c}}(r_c^2-2Q^2)(Q^2+r_c^2-2r_c), \quad \Omega_c^2=\frac{1}{r_c^4}(r_c^2-2r_c+Q^2),
\end{equation}

that substituting $r_c$ from (\ref{rcRN}) give

\begin{equation}
\lambda^2= \frac{16 \sqrt{9-8Q^2}(3-2Q^2+ \sqrt{9-8Q^2})}{(3+ \sqrt{9-8Q^2})^5}, \quad \Omega_c^2=\frac{8(3-2Q^2+ \sqrt{9-8Q^2})}{(3+ \sqrt{9-8Q^2})^4}.
\end{equation}

These are the expressions that we will use in the comparisons with the NLEM cases.
In the RN case analytic solutions can be found all the way through. In none of the NLEM examples presented analytic solutions were determined.
Recent works on aspects of RN-QNM are \cite{Richartz:2014jla} \cite{Corda:2013paa}.

\subsection{Born-Infeld black hole}

 The BI nonlinear electrodynamics was first considered by Born and Infeld  in an attempt to cure at the classical level the singularity of the electric field of a point charge \cite{Born:1934gh}.
Born-Infeld electrodynamics possesses several interesting physical features like the absence of birrefringence,  the Maxwell limit for weak electromagnetic field, as well as the finiteness of the electric field at the charge position. The Einstein$-$Born$-$Infeld (EBI) generalization of the Reissner Nordstr\"{o}m (RN) black hole
was obtained by Garc\'ia, Salazar and Pleba\~nski  in \cite{GarcíaD.1984}; the geodesic structure of BI black hole was studied in \cite{Breton:2002td} and recently in \cite{Linares:2014nda}.This Einstein$-$Born$-$Infeld solution is singular at the origin and it is characterized by three parameters: mass, charge and the BI parameter; the BI black hole  can present one, two or none horizons depending on the values of the parameters. The Born$-$Infeld Lagrangian is given by

\begin{equation}(\label{BIlagr}
L(F)=4b^{2}\left(-1+\sqrt{1+\frac{F}{2b^{2}}}\right).
\end{equation}

For a line element of the form (\ref{sss}) the solution to the EBI equations  is given by the metric function
\begin{equation}
f(r)=g(r) = 1-\frac{2M}{r}+\frac{2}{3}r^{2}b^{2}\left(1-\sqrt{1+\frac{q^{2}}{b^{2}r^{4}}}\right)+\frac{2}{3}\frac{q^{2}}{r}\sqrt{\frac{b}{q}}
\mathbb{F}\left(\arccos\left[\frac{br^{2}/q-1}{br^{2}/q+1}\right],\frac{1}{\sqrt{2}}\right),
\end{equation}
where $\mathbb{F}$ is the elliptic integral of the first kind, $M$ is the mass parameter, $q$ is the electric charge (both in length units) and $b$ is the Born-Infeld parameter that corresponds to the magnitude of the electric field at $r=0$. The nonvanishing component of the electromagnetic field is $F_{rt}=q\left(r^{4}+\frac{q^{2}}{b^{2}}\right)^{-1/2}$. The nonlinear factors $G_m$ and $G_e$ are, in case that we are considering only electric charge, $q_m=0$,

 \begin{equation}
G_{m}=1, \quad G_{e}=\left(1+\frac{q^{2}}{b^{2}r_c^{4}}\right).
\end{equation}

\begin{figure}[h]\label{BIW}
\begin{center}
\includegraphics [width =0.42 \textwidth ]{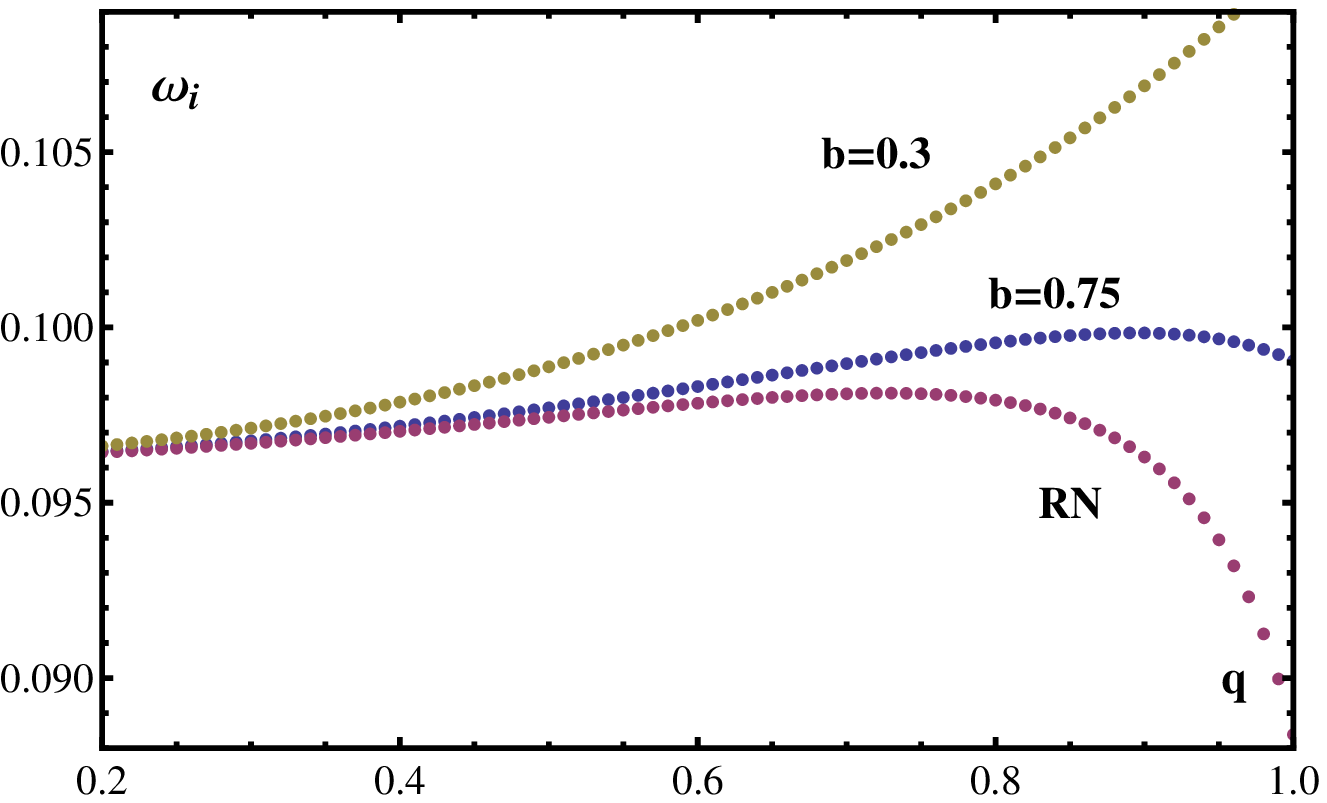}
\includegraphics [width =0.42 \textwidth ]{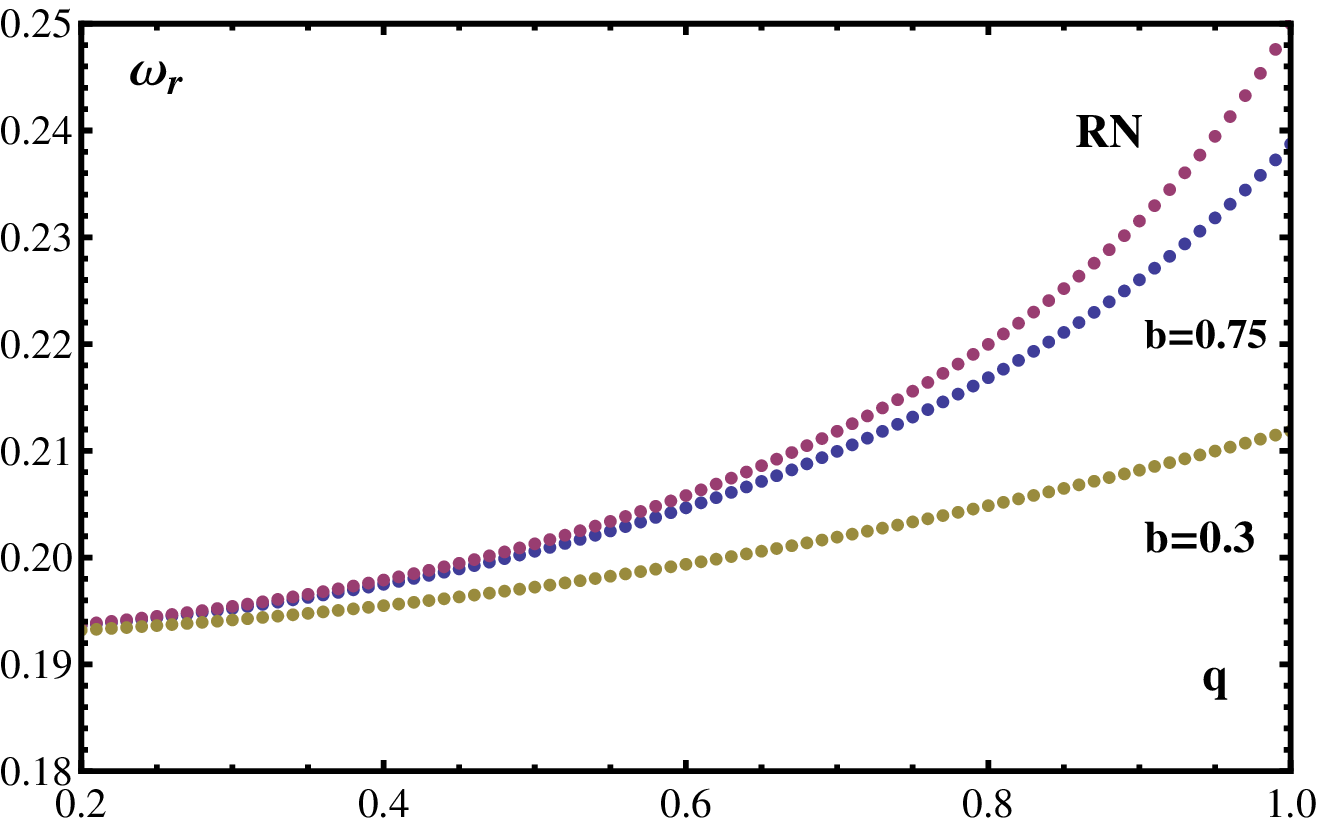}
\end{center}
\caption{The behavior of  $\omega_{i}$ and $\omega_{r}$ For the BI and RN black holes is shown as functions of the charge. Two values of $b$ were considered, $b=0.3$ and $b=0.75$; the other parameters fixed: $M=1$ and $n=0$. }\label{BIW}
\end{figure}

Hereafter we consider $\omega_r \mapsto \omega_r/l$ for our analysis  \cite{Fernando:2012yw}.
In Fig. \ref{BIW} it is  shown the behavior of the QNFs  $\omega_{i}$ and $\omega_{r}$  of the BI black hole for two different values of the parameter $b$ and are compared with RN black hole.
The imaginary part of QNF $\omega_{i}$,  for RN case, increases as $q$ augments and presents a maximum then decreasing; for RN the value of $q$ cannot exceed $q=1$ that corresponds to the extreme black hole, $q=M$. BI does not have this constraint, and
$\omega_{i}$ increases without bound for small values of $b$.  Therefore
BI-$\omega_{i}$ is enhanced as compared with RN. The opposite occurs with the BI-$\omega_{r}$ that is suppressed as compared with RN. Both frequencies approach the RN limit as $b$ increases.

The QNF  coming from  massless particles and photons (ph)  are compared in Fig. \ref{BIWC}. For the imaginary part, $\omega_{i}$, the previous behavior is enhanced even more in photon trajectories, while $\omega_{r}$-massless particles values are greater that photon's.
Massless test particles QNFs were analyzed in \cite{Fernando:2004pc}  using the WKB approximation.

\begin{figure}[h]\label{BIWC}
\begin{center}
\includegraphics [width =0.42 \textwidth ]{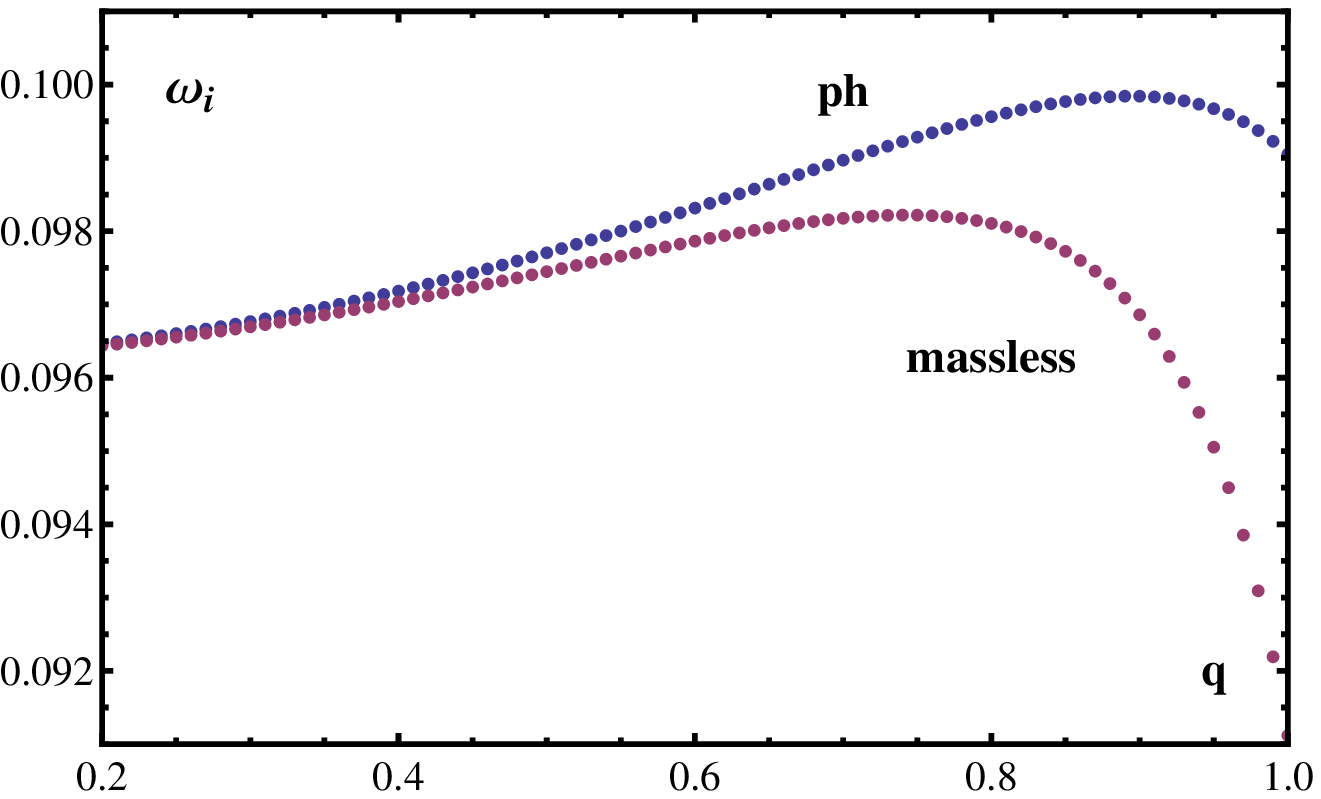}
\includegraphics [width =0.42 \textwidth ]{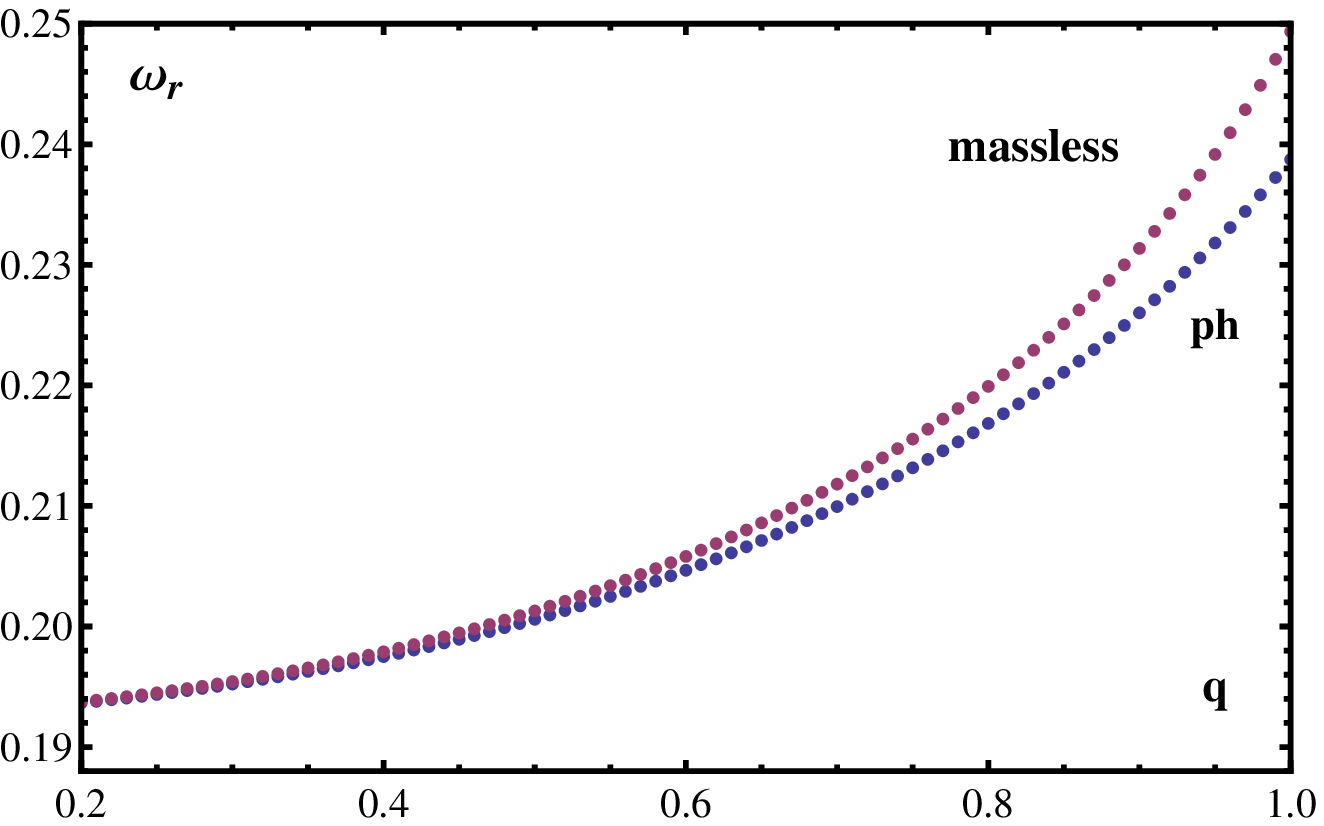}
\end{center}
\caption{Comparison between $\omega_{i}$ and $\omega_{r}$ of the BI black hole for massless particles and photon (ph) trajectories is shown. The rest of parameters are $M=1$, $b=0.75$  and $n=0.$ }\label{BIWC}
\end{figure}

\subsection{Euler-Heisenberg Black hole}

The effective action for electrodynamics due to one-loop quantum corrections was calculated by Euler and  Heisenberg (EH). For the low-frequency limit $\omega \ll m_{e}c^{2}/h^{2}$ the effective Lagrangian with magnetic charge \cite{Yajima:2000kw} takes the form

\begin{equation}\label{EHlagr}
L(F)= F(1-aF)
\end{equation}

with $a=he^{2}/(360\pi^{2}m_{e}^{2})$, where $h$, $e$, and $m_{e}$ are the Planck constant, electron charge, and electron
mass, respectively.  BI theory applies for fields even stronger than the ones in QED, because it turns out that a Lagrangian very similar to  (\ref{EHlagr}) can be obtained in the ``weak" field limit of the BI Lagrangian:
expanding the square root in  (\ref{BIlagr})  up to second order, $(1+x)^{1/2}=1+x/2-x^2/8+ \cdots$,  we obtain

\begin{equation}
\label{BIlagr_exp}
L(F)=4b^{2}\left(-1+ 1+\frac{F}{4b^{2}} + \frac{F^2}{32b^{4}} +\cdots \right) = F - \frac{F^2}{8b^{2}},
\end{equation}

from the comparison  with (\ref{EHlagr})  a relationship between the EH and BI parameters  is obtained, $8b^2=a^{-1}$.
For the previous reasons we expect for the EH-QNFs a very similar behavior than the BI frequencies. Indeed this is the case.

In  \cite{Yajima:2000kw} a solution of the  EH field coupled to gravity equations was found (see also \cite{DeLorenci:2001bd}). It corresponds to a SSS magnetic black hole with metric  elements in  (\ref{sss}) given by

\begin{equation}
f(r)=g(r)=1-\frac{2M}{r}+\frac{q^{2}}{r^{2}}-\frac{2}{5}a\frac{q^{4}}{r^{6}}
\end{equation}

where
$M$ is the mass parameter, $q=q_m$ is the magnetic charge.  One or two horizons may occur: if  $(M/q)^{2}\leq 24/25$ a single horizon exists but for  $(M/q)^{2}> 24/25$ a second and a third horizons occur. Extremal solutions exist only for $a^{2}\leq a^{2}_{crit}=(8/27)q^{2}$. The electromagnetic invariant is  $F={2q^{2}}/{r^{4}}$. Since $q_e=0$,   the $G_e$ and $G_m$ factors are

\begin{equation}
G_{e}=1, \quad G_{m}=\left(1+\frac{8aq^{2}}{4aq^{2}-r_c^{4}}\right)
\end{equation}

\begin{figure}[h]\label{EHW}
\begin{center}
\includegraphics [width =0.42 \textwidth ]{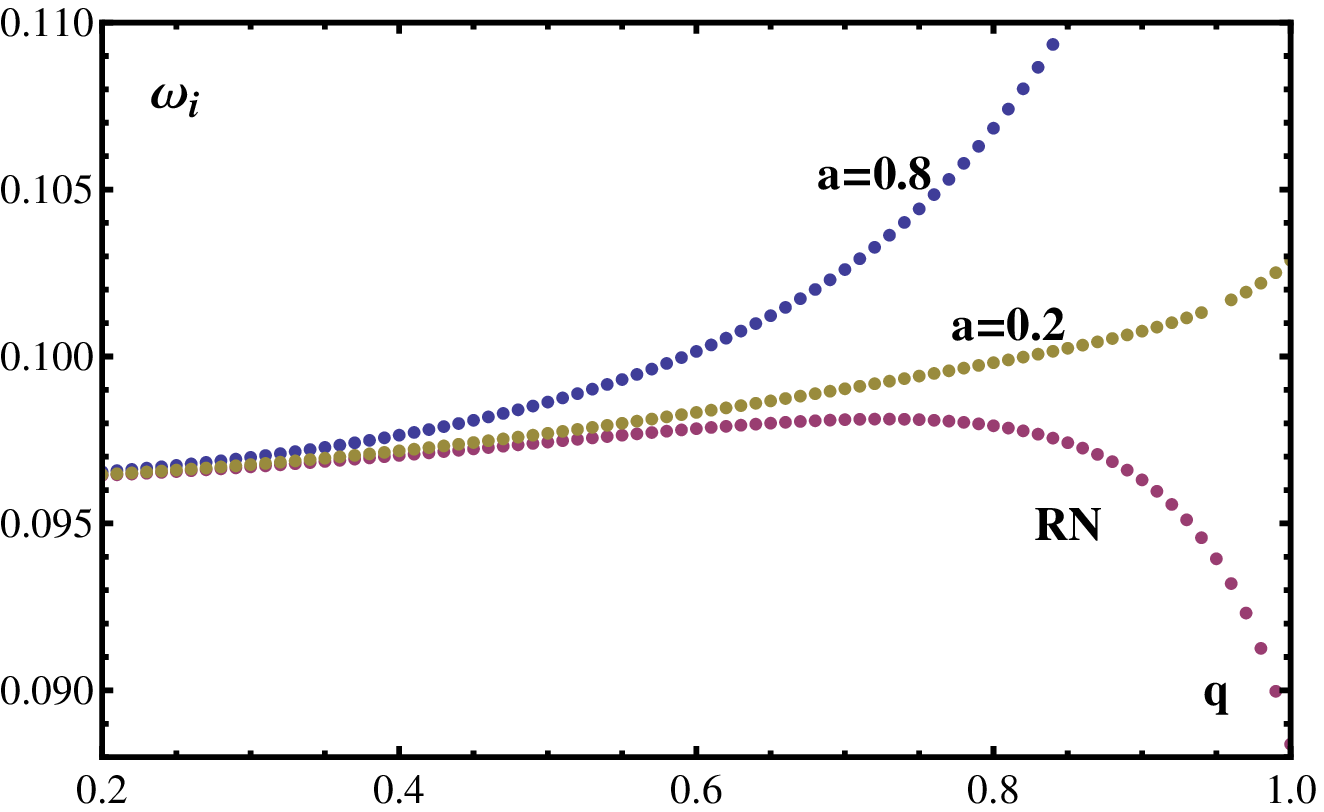}
\includegraphics [width =0.42 \textwidth ]{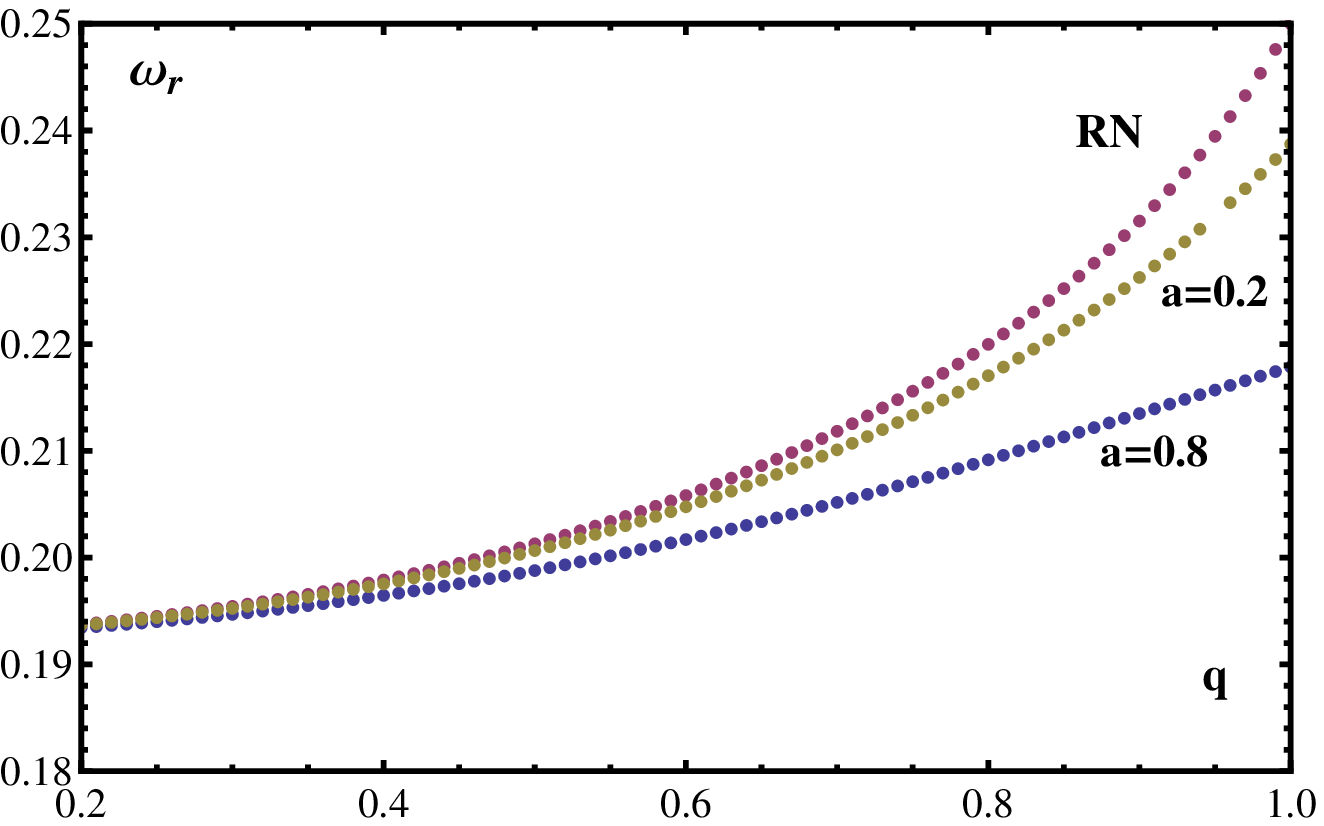}
\end{center}
\caption{The behavior of  $\omega_{i}$ and $\omega_{r}$ in terms of the charge $q$ is shown for the EH and RN black holes; the rest of parameters fixed as $M=1$ and $n=0$}\label{EHW}
\end{figure}

The behavior of QNFs $\omega_{i}$ and $\omega_{r}$ for EH black hole resembles the BI ones (see Fig. \ref{EHW}), taking into account that the nonlinear parameters $b$ and $a$ are inversely proportional, for small $a$'s we obtain an effect similar to
large $b$'s, i.e. the behavior approaches the RN one. While for larger $a$'s the departure from RN behavior is clear, enhanced for $\omega_{i}$ and suppressed for $\omega_{r}.$ Regarding the massless-photon comparison, $\omega_{i}$ is enhanced and $\omega_{r}$ is suppressed for photons respect to massless particle trajectories as can be seen in Fig.\ref{EHWC}.

\begin{figure}[h]\label{EHWC}
\begin{center}
\includegraphics [width =0.42 \textwidth ]{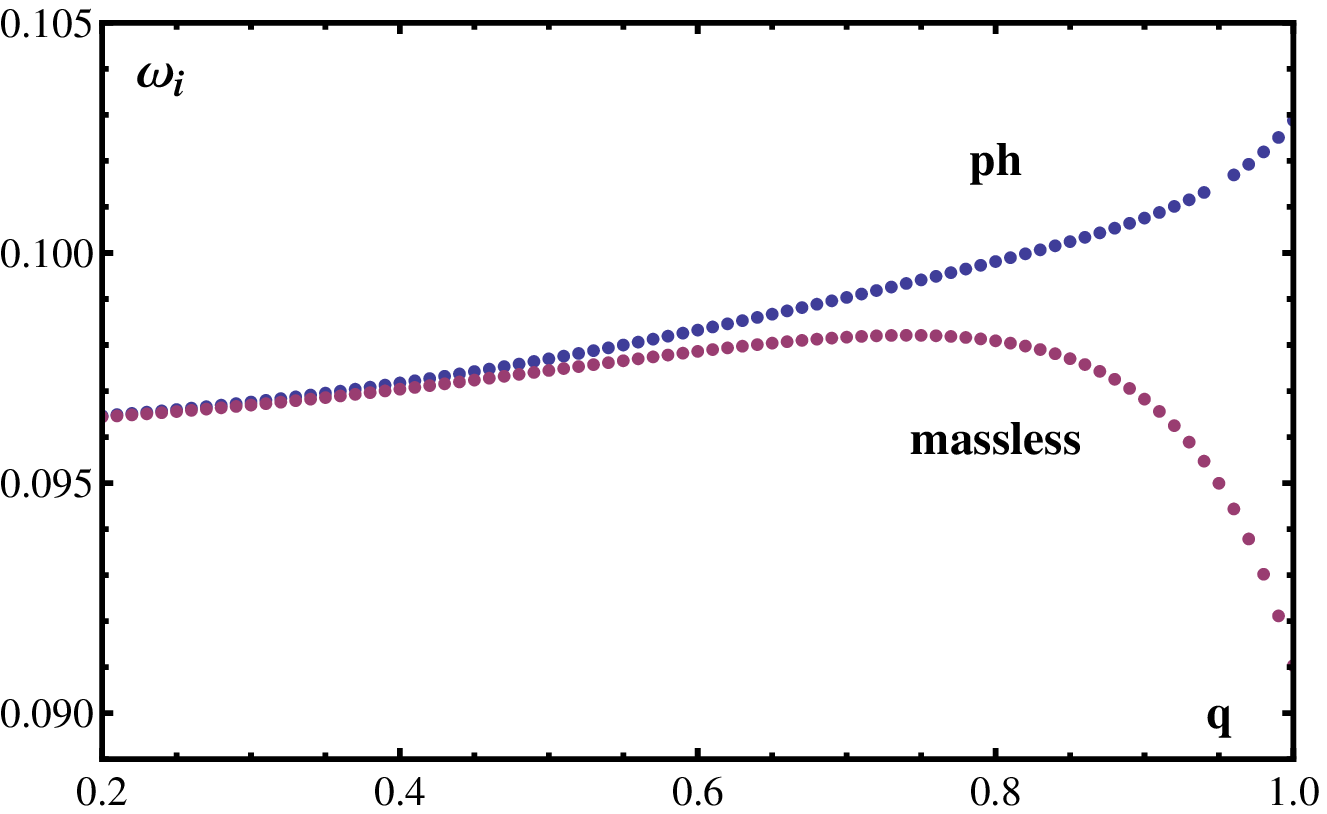}
\includegraphics [width =0.42 \textwidth ]{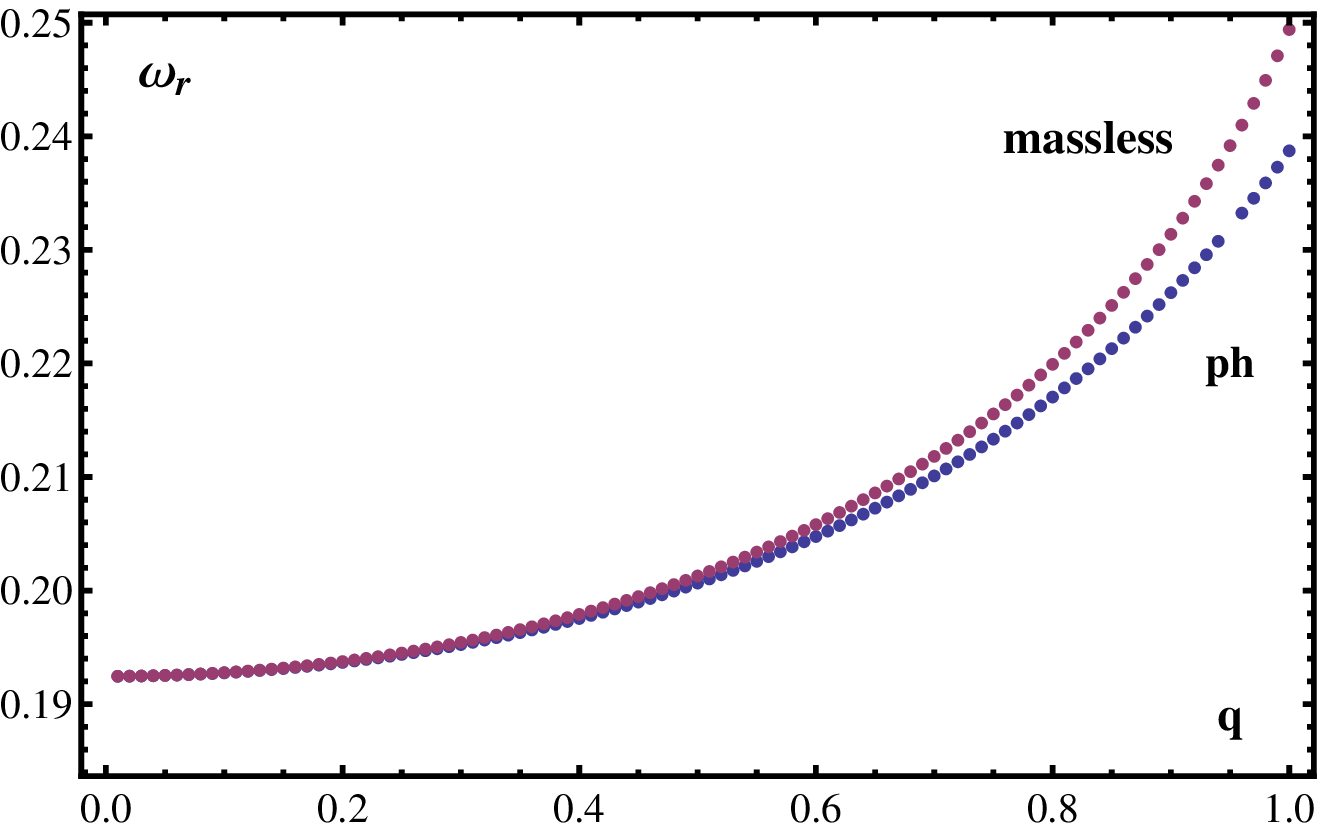}
\end{center}
\caption{Comparison for varying charge of photon vs. massless particle QNFs, imaginary and real parts, $\omega_{i}$ and $\omega_{r}$, for the EH black hole. In this plot $M=1$, $a=0.2$  and $n=0$}\label{EHWC}
\end{figure}

\subsection{Bardeen black hole}

This  model was proposed by Bardeen in the sixties as an example of a regular  black hole; later on  Ay\'on-Beato and Garc\'ia  \cite{AyonBeato:2000zs} found a nonlinear electromagnetic source for such a black hole. Hence, the
Bardeen black hole can be interpreted as a self-gravitating  nonlinear magnetic monopole with mass $M$ and charge
$q_{m}=g$,  derived from  Einstein gravity coupled to the Lagrangian

\begin{equation}
L(F)=\frac{6}{s g^{2}}\frac{(g^{2}F/2)^{\frac{5}{4}}}{(1+\sqrt{g^{2}F/2})^{\frac{5}{2}}}
\end{equation}
where  $s= |g|/2M$.
A solution for the coupled Einstein-Bardeen Lagrangian for a static spherically symmetric space has the metric function $f(r)=g(r)$ given by

\begin{equation}\label{Bardeenmetric}
f(r)=1-\frac{2Mr^{2}}{(r^{2}+g^{2})^{\frac{3}{2}}},
\end{equation}

The solution (\ref{Bardeenmetric}) presents horizons only if $2s= g/m \leq 0.7698.$  The electromagnetic invariant is $F=2g^{2}/r^{4}$ and the $G_e$ and $G_m$ factors are  given by

\begin{equation}
G_{e}=1, \quad G_{m}=1-\frac{4(6g^{2}-r^{2})}{8(r^{2}+g^{2})}.
\end{equation}

The radius of the unstable circular orbit
$r_c$ is found numerically from (\ref{rcEq.}) and care must be taken in that such root be greater than the horizon radius, $r_c>r_h$.

The behavior of  the QNMs from light rays impinging upon Bardeen black hole are shown in Fig.\ref{BAW}.  $\omega_{i}$ decreases when $g$ increases and similarly for massless  particles and photons. This behavior is in contrast with the two previously analyzed examples; the resemblance with RN $\omega_{i}$ occurs only when the charge approaches zero. Regarding $\omega_{r}$ the behavior for massless particles is indistinguisable from  RN one. The one corresponding to photons is odd in the sense that not even in the limit that charge goes to zero the RN behavior is recovered; one possible explanation may be the nature of the solution because charge and mass parameters are not independent, and in fact when charge is turned off, so does the mass whose origin is purely electromagnetic. QNFs for massless particles have been determined in \cite{Fernando:2012yw}.

\begin{figure}[h]\label{BAW}
\begin{center}
\includegraphics [width =0.42 \textwidth ]{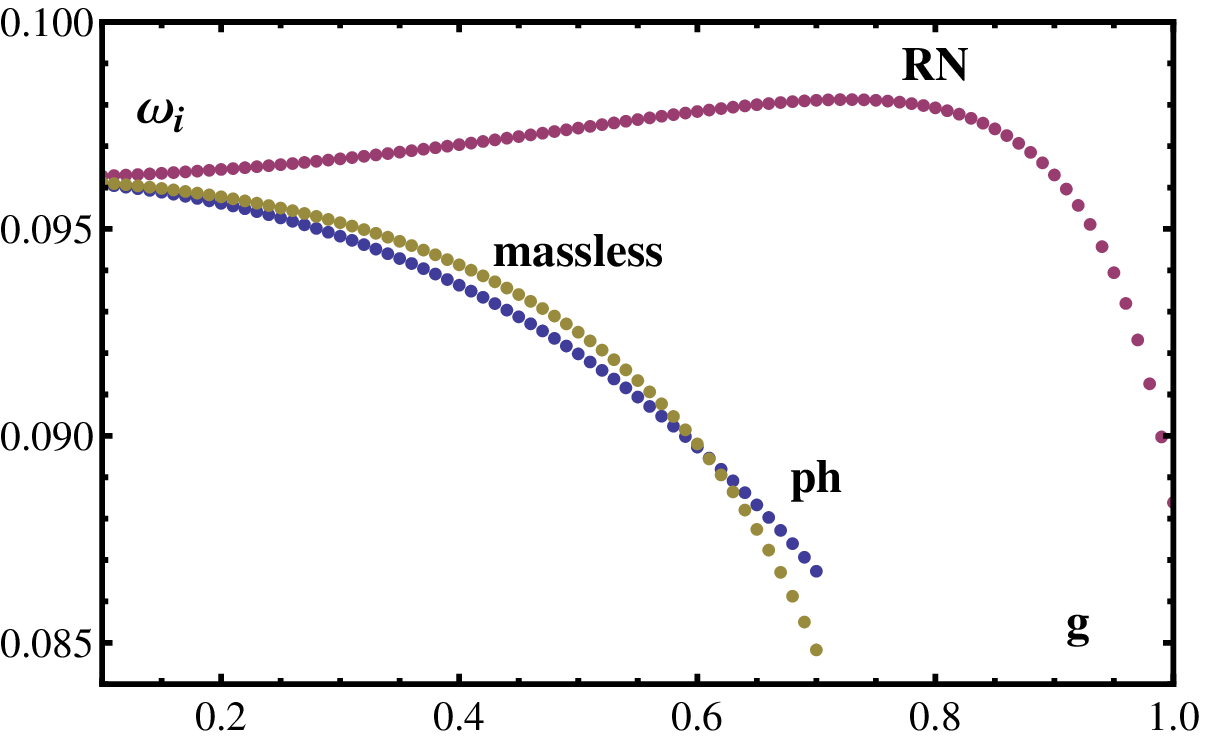}
\includegraphics [width =0.42 \textwidth ]{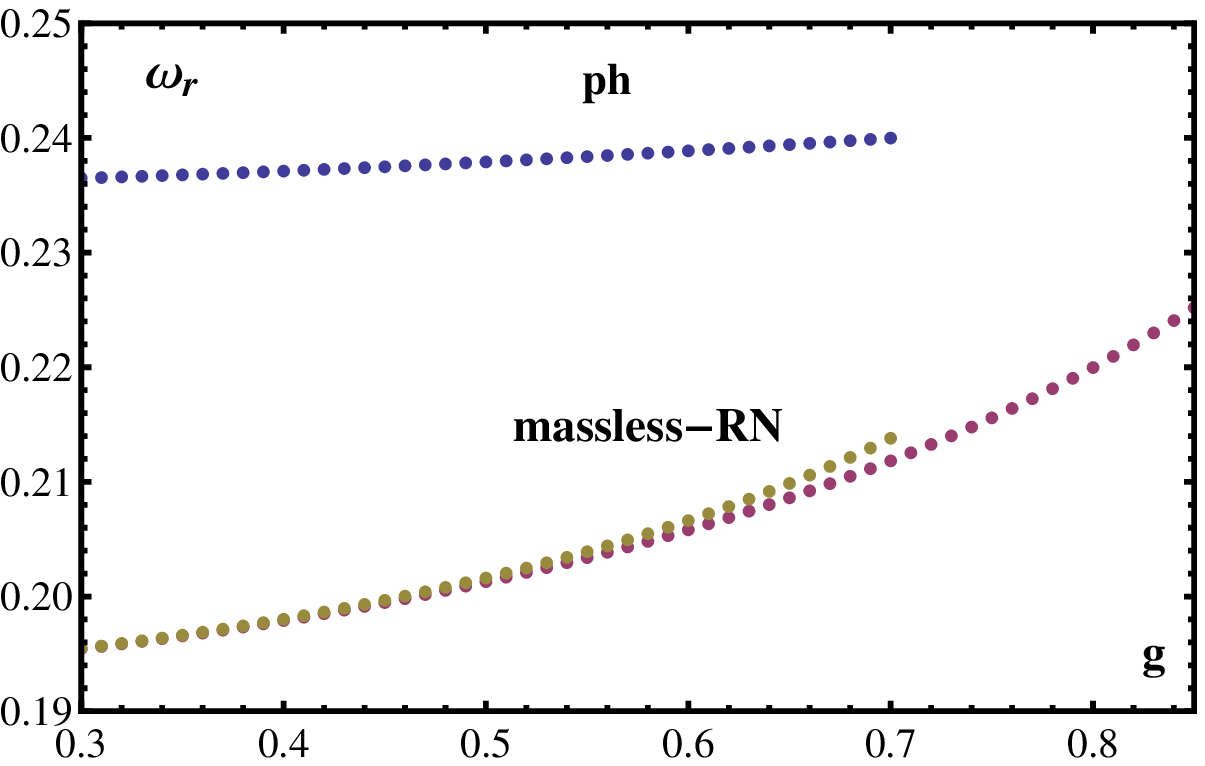}
\end{center}
\caption{It is shown the behavior of  $\omega_{i}$ and $\omega_{r}$ of the Bardeen black hole  as  functions of the charge $g$, keeping fixed $M=1$ and $n=0.$}\label{BAW}
\end{figure}

\subsection{Bronnikov magnetic black hole}

In \cite{Bronnikov:2000vy}   NLEM Lagrangians coupled to gravity  were  analyzed focusing on the properties that lead to nontrivial regular metrics. No go theorems forbid regular electric black holes,  but magnetic ones can be found. One example is the presented in the same reference,
with the Lagrangian

\begin{equation}
L(F)=F {\rm sech}^2 [a(F/2)^{1/4}]
\end{equation}
where $a$ is a constant. The metric function in the line element of the form (\ref{sss}) is

\begin{equation}\label{Bronnokovmetric}
f(r)= 1-\frac{g^{3/2}}{ar} \left[ {1- \tanh \left( \frac{a \sqrt{g}}{r}\right)} \right]
\end{equation}

the constant $a$ is related to the mass $m$ and the magnetic charge $g$ by $a=g^{3/2}/(2M)$. The solution corresponds to a black hole if $M/g > 0.96$. The electromagnetic invariant  is $F=2g^{2}/r^{4}$. For a purely magnetic charge $G_{e}=1$ and

\begin{equation}
 G_{m}=\frac{g^{2}\sinh^{2}(\frac{g^{2}}{2Mr})\left(-2g^{2}+g^{2}\cosh(\frac{g^{2}}{2Mr})-5Mr\sinh(\frac{g^{2}}{2Mr})\right)}{4Mr\left(-4Mr^{3}+g^{2}\tanh(\frac{g^{2}}{2Mr})\right)}
\end{equation}

The QNFs behavior when the charge is varying is shown in Fig. \ref{BROW}. The imaginary part of QNF $\omega_{i}$ for  massless particles as well as for photon reproduces the RN one, and slight departure is observed only when the charge approaches its upper limit. As for the real QNF, $\omega_{r}$, for massless particles  is coincident with RN for any charge, while small difference occurs for photons. From all of the examined examples Bronnikov black hole is the one in which the nonlinear effects are the least.

\begin{figure}[h]\label{BROW}
\begin{center}
\includegraphics [width =0.42 \textwidth ]{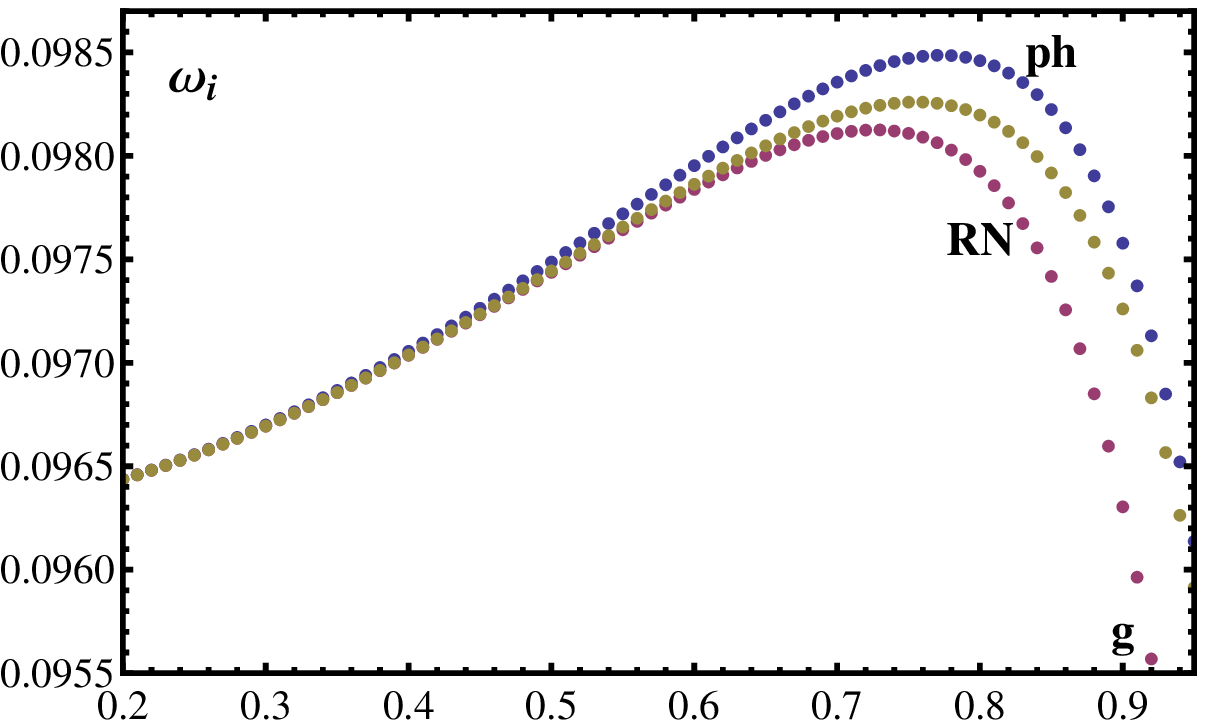}
\includegraphics [width =0.42 \textwidth ]{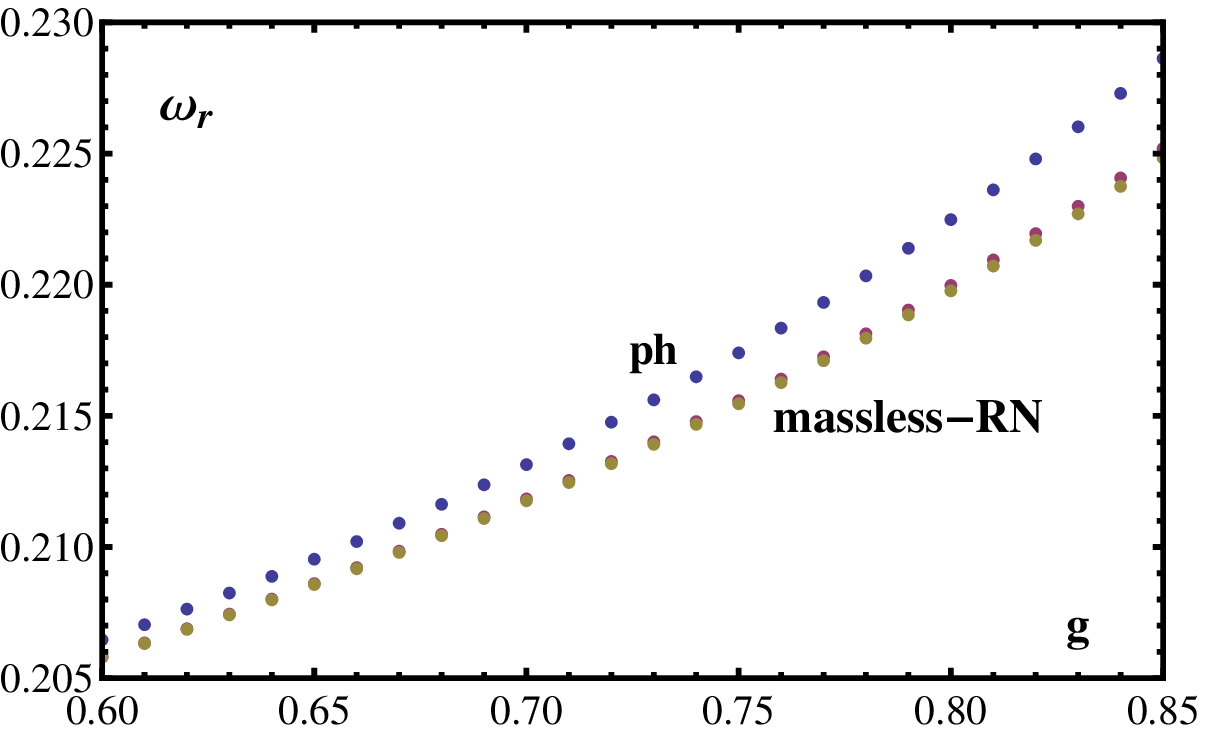}
\end{center}
\caption{QNFs  $\omega_{i}$ and $\omega_{r}$ of the Bronnikov black hole  are shown as functions of the charge $g$; the other parameters fixed to $M=1$ and $n=0.$}\label{BROW}
\end{figure}

\section{Conclusions}

We have studied the QNM frequencies of NLEM black holes through Lyapunov exponent in the optical approximation. QNFs were calculated from the unstable null geodesics of the effective metric. Effective metric is obtained from the regular metric but taking into account the NLEM effects.
From the expressions of the real and imaginary parts of the QNFs and Eqs.  (\ref{expLyapunov1}) and (\ref{angular2}), it is clear that the NLEM effects will modify QNF enhancing or suppressing them, depending on the sign of the quotients  $L_{FF}/L_{F}$ and of the second derivative of $G_{e}/G_{m}$.
The NLEM effects manifest on  the dynamics of the light perturbations modifying the oscillation periods as well as the damping times.

The obtained QNFs, in the eikonal approximation,   are calculated to four SSS black hole solutions corresponding to different NLEM theories coupled to gravity. In all cases comparison is done with the QNFs of  the  linear counterpart, the RN back hole. In three of the examples (BI, EH and Bronnikov solution), the NLEM effect when the charge is varied consists in suppressing the real part while increasing the imaginary one, i.e. oscillation periods are larger and relaxation occurs faster.

Comment apart deserves the Bardeen black hole. In this case the effect is the opposite: the real QNF is enhanced while the imaginary part is suppressed; notoriously the real frequency is not recovered when the charge approaches zero. Several explanations come to mind, among them is that mass and charge parameters are not independent in this solution and the limit to zero of the charge cannot be taken separatedly from the value of the mass, since the origin of the mass is magnetic; this is a self-gravitating magnetic structure.  Additionaly, the comparison between the behavior of massless particles and photons was done.

A thorough analysis is needed to know how the modifications introduced by NLEM effects influence the black hole  stability.  Another point of interest would be checking if there is any consequence of introducing NLEM related to the conjectured relationship between the real part of the QNF and the black hole area quantization.


\begin{acknowledgments}
N. B. acknowledges partial support of CONACyT (Mexico), project 166581
\end{acknowledgments}

\bibliographystyle{plain}
\bibliography{bibliografia}

\end{document}